\documentclass[conference]{IEEEtran}
\usepackage[buvect]{commands09}
\usepackage{vect09,matrix09}
\usepackage{ushort}
\usepackage{verbatim}
\usepackage{cite}
\usepackage{setspace}
\usepackage{graphicx}
\usepackage{amsfonts, amsmath, amssymb}
\usepackage{algorithm}
\usepackage{algorithmic}
\usepackage{paralist}

\allowdisplaybreaks

\DeclareMathOperator{\thevee}{\vee}

\newtheorem{proposition}{Proposition}

\title{Rapid-On-Off-Division Duplex for Mobile
Ad Hoc Networks\thanks{This work was supported by DARPA under grant
   W911NF-07-1-0028.}}

\title{Virtual Full-Duplex Wireless Communication via Rapid On-Off-Division Duplex\thanks{This work was supported by DARPA under grant
   W911NF-07-1-0028.}}

\author{\IEEEauthorblockN{Dongning Guo and Lei Zhang}
 \IEEEauthorblockA{Department of Electrical Engineering \& Computer Science \\
    Northwestern University \\
    Evanston, IL 60208, USA} }

\IEEEoverridecommandlockouts

\begin{document}
\pagestyle{empty}

\maketitle
\thispagestyle{empty}
\begin{abstract}
  This paper introduces a novel paradigm for designing the physical
  and medium access control (MAC) layers of mobile ad hoc or
  peer-to-peer networks formed by half-duplex radios.  A node equipped
  with such a radio cannot simultaneously transmit and receive useful
  signals at the same frequency.  Unlike in conventional designs,
  where a node's transmission frames are scheduled away from its
  reception, each node transmits its signal through a randomly
  generated on-off {\em duplex mask (or signature)} over every frame
  interval, and receive a signal through each of its own off-slots.
  This is called {\em rapid on-off-division duplex (RODD)}.  Over the
  period of a single frame, every node can transmit a message to some
  or all of its peers, and may simultaneously receive a message from
  each peer.  Thus RODD achieves virtual full-duplex communication
  using half-duplex radios and can simplify the design of higher
  layers of a network protocol stack significantly.
  The throughput of RODD is evaluated under some general settings,
  which is significantly larger than that of ALOHA.  RODD is
  especially efficient in case the dominant traffic is simultaneous broadcast
  from nodes to their one-hop peers, such as in spontaneous wireless
  social networks, emergency situations or on battlefield.
  Important design issues of peer discovery, distribution of on-off
  signatures, synchronization and error-control coding are also
  addressed.
\end{abstract}



\section{Introduction}\label{s:int}

In spite of decades of advances in wireless and networking
technologies, to design a functional and reliable mobile ad hoc or
peer-to-peer network remains enormously challenging~\cite{AndJin08CM}.
The main roadblocks include the difficult nature of the wireless
medium and the mobility of wireless terminals, among others.
A crucial constraint on wireless systems is the {\em half-duplex} nature
of affordable radios, which prevents a radio from receiving any useful signal
at the same time and over the same frequency band within which it is
transmitting.  The physical reason is that during
transmission, a radio's own signal picked up by its receive antenna is
typically orders of magnitude stronger than the signals
from its peers, such that the desired signals are lost
due to the limited dynamic range of the radio frequency (RF)
circuits.  The half-duplex constraint has far-reaching consequences in
the design of wireless networks:
The uplink and downlink transmissions in any cellular-type network are
separated using
time-division duplex (TDD) or
frequency-division duplex (FDD); standard designs of wireless ad hoc
networks schedule transmission frames of a node away from the time and
frequency slot over which the node receives data~\cite{XuSaa01CM}.

In this work, the half-duplex constraint is addressed at a
fundamental level, which is that the received signal of a half-duplex
node is {\em erased} during periods of its own active transmission.
We recognize that, it is neither necessary nor efficient to separate the
transmission slots and listening slots of a node in the timescale of a
frame of hundreds or thousands of symbols as in TDD. 
We propose for the first time a technique called {\em rapid
  on-off-division duplex (RODD)}.
The key idea is to let each node transmit according to a unique {\em
  on-off duplex mask (or signature)} over a frame of symbols or
slots, so that the node can receive useful signals from its peers
during the {\em off-slots} interleaved between its {\em on-slot} transmissions.
Importantly, all nodes may send (error-control-coded) information
simultaneously over a frame interval, as long as the masks of peers
are sufficiently different, so that a node receives enough signals
during its off-slots to decode information from its peers.  Over
the period of a single frame, every node simultaneously broadcasts a
message to some or all other nodes in its neighborhood, and may receive a
message from every neighbor at the same time.

Switching the carrier on and off at the timescale of one or several
symbols is feasible, thanks to the sub-nanosecond response time of RF
circuits.  In fact, on-off signaling over sub-millisecond slots is
used by time-division multiple-access (TDMA) cellular systems such as
GSM.  Time-hopping impulse radio transmits on and off at nanosecond
intervals~\cite{WinSch00TC}, which is orders of magnitude faster than 
needed by RODD (in microseconds). 
Moreover, receiving signals during one's own off-slots avoids
self-interference and circumvents the dynamic range issue which
plagues other full-duplex schemes, such as code-division duplex
(CDD)~\cite{AsaSat96IEICE, Lee02CM}.

Ad hoc networks using rapid on-off-division duplex have unique
advantages:
\begin{inparaenum}
\item RODD enables virtual full-duplex transmission and greatly
  simplifies the design of higher-layer protocols.  In particular,
  ``scheduling'' is carried out in a microscopic timescale over the
  slots, so that there is no need to separate transmitting and
  listening frames;
\item RODD signaling takes full advantage of the superposition and
  broadcast nature of the wireless medium.  As we shall see, the
  throughput of a RODD-based network is greater than that of
  ALOHA-type random access, and is more than twice as large as that of
  slotted ALOHA in many cases;
\item RODD signaling is particularly efficient when the traffic is
  predominantly peer-to-peer broadcast, such as in mobile systems used
  in  local advertising, spontaneous social networks, emergency
  situations or on battlefield;
\item Communication overhead usually comes as an afterthought in
  network design, whereas RODD enables extremely efficient exchange of
  a small amount of state information amongst neighbors;
\item Because nodes simultaneously transmit, the channel-access delay
  is typically smaller and more stable than in conventional
  reservation or scheduling schemes. 
\end{inparaenum}

This paper presents a preliminary study of several aspects of RODD.
Related work and technologies are discussed in
Section~\ref{s:related}.  Mathematical models of a network of nodes
with synchronous RODD signaling is presented in Sections~\ref{s:model}.  Assuming
mutual broadcast traffic, the throughput of a fully-connected,
synchronized, 
RODD-based network is studied in Section~\ref{s:capacity}.  Design
issues such as duplex mask assignment, peer discovery, error-control
codes and synchronization are discussed in
Sections~\ref{s:peer}--\ref{s:sync}.  We conclude the paper with a
discussion of applications of RODD in Section~\ref{s:con}.

\section{Related Work}
\label{s:related}

There have been numerous works on the design of physical and MAC
layers for wireless networks (see the surveys~\cite{ShaRap03CM, 
 KumRag06AdHocNet, RubMor06IFIP} and references therein). 
Two major challenges need to be addressed: One is the half-duplex
constraint;
the other 
is the broadcast and superposition nature of the wireless medium, so
that simultaneous transmissions interfere with each other at a receiver. 
State-of-the-art designs either schedule nodes orthogonally ahead of
transmissions, or apply an ALOHA-type random access scheme, or use a
mixture of random access and scheduling reservation~\cite{MohKri05}. 
Typically, the {\em collision model}\, is assumed, where if multiple
nodes simultaneously transmit, their transmissions
fail due to collision at the receiver.
Under such a model, random access leads to poor efficiency (e.g., ALOHA's
efficiency is less than $1/e$).  On the other hand, scheduling
node transmissions is often
difficult and subject to the hidden terminal and exposed terminal
problems. 


Despite the half-duplex
constraint, it is neither necessary nor efficient to separate a node's
transmission slots and listening slots in the timescale of a frame.
In fact, time-sharing can fall considerably short of the theoretical
optimum.
For example, it has been shown that the capacity of a cascade of two
noiseless binary bit pipes through a half-duplex relay is 1.14 
bits per channel use~\cite{LutHau08ISIT, LutKra10ISIT}, which far exceeds
the 0.5 bit achieved by TDD and even the 1 bit upper bound on the rate
of binary signaling. 
This is because non-transmission can be regarded as an additional
symbol for signaling (besides 0 and 1), whose positions can be used
to communicate information (see also~\cite{Kramer07}).

Several recent works on the implementation of physical and MAC layers
break away from the collision model and single-user transmission.  For
example, superposition coding for degraded broadcast channels has been
implemented using software-defined radios~\cite{GanGon10ICC}.
Analog network coding has also been implemented based on
802.11 technology~\cite{KatGol07Sigcomm}, where, 
when two senders transmit
simultaneously, their packets collide, or more precisely, superpose at
the receiver, so that if the receiver already knows the content of one of the
packets, it can cancel the interference and decode the other packet.
Similar ideas have been proven feasible in some other contexts to achieve
interference cancellation in unmanaged
ZigBee networks~\cite{HalAnd08Mobicom}, ZigZag decoding for 802.11
in~\cite{GolKat08Sigcomm}, and interference alignment and cancellation
in~\cite{GolPer09Sigcomm}.



Rapid on-off-division duplex is related to code-division duplex, which
was proposed in the context of code-division multiple access
(CDMA)~\cite{AsaSat96IEICE}.  In CDD, orthogonal (typically antipodal)
spreading sequences are allocated to uplink and downlink
communications, so that a receiver ideally cancels self-interference by
matched filtering with its own receive spreading sequence.  Despite
the claimed higher spectral efficiency than that of TDD and FDD
in~\cite{Lee02CM}, CDD is not used in practice because it is difficult
to maintain orthogonality due to channel impairments and suppress
self-interference which is orders of magnitude stronger than the
desired signal.  
In RODD, the desired signal is sifted through the off-slots of the
transmission frame, so that the leakage of the transmit energy into
the received signal is kept to the minimum.  

RODD can also be viewed as (very fast) TDD with irregular symbol-level  
transition between transmit and receive slots as well as coding over  
many slots.
Although 
on-off signaling can in principle be applied to the frequency domain, 
it would be much harder to implement sharp band-pass filters to remove 
self-interference.

The RODD signaling also has some similarities to  that of
time-hopping impulse radio~\cite{Scholt93Milcom, WinSch98CL}.
Both schemes transmit a sequence of randomly spaced pulses.  There are
crucial differences: Each on-slot (or pulse) in RODD spans
one or a few data symbols (in microseconds), whereas each pulse in
impulse radio is a baseband monocycle of a nanosecond or so duration.  
Moreover, impulse radio is carrier-free and spreads the spectrum by
many orders of magnitude, whereas RODD uses a carrier and is not
necessarily spread-spectrum.

\section{Models and RODD Signaling}
\label{s:model}

Consider an ad hoc network consisting of $K$ nodes, indexed by
$1,\dots,K$.  Suppose all transmissions are over the same frequency
band.
Suppose for simplicity each slot is of one symbol interval and
all nodes are perfectly
synchronized over each frame of $M$ slots.  Let the binary on-off
duplex mask of node $k$ over slots~$1$ through $M$
be denoted by $\s_k=[s_{k1},\dots, s_{kM}]$.
During slot $m$, node $k$ may transmit a symbol
if $s_{km}=1$, whereas if $s_{km}=0$, the node listens to the channel
and emits no energy.

\subsection{The Fading Channel Model}

The physical link between any pair of nodes is modeled as a fading
channel.  Let the path loss satisfy a power law with exponent $\alpha$.
The received signal of node $k$ during each slot $m\in\{1,\dots,M\}$
is described by
\begin{equation} \label{eq:MAC} %
 Y_{km} = (1-s_{km}) \sum_{j\ne k} \sqrt{\gamma_j} d_{kj}^{-\alpha/2}
 h_{kj} s_{jm} X_{jm} + W_{km}
\end{equation}
where $d_{kj}$ denotes the distance between nodes $k$ and $j$,
$h_{kj}$ denotes the fading coefficient, $X_{jm}$ denotes the
transmitted symbol of node $j$ at time slot $m$, $W_{km}$ denotes
additive noise, and $\gamma_j$ denotes the signal-to-noise ratio (SNR)
of node $j$ at unit distance without fading.
The received signal of each node over its own off-slots is
the superposition of the signals from its peers over those slots (in
addition to noise).  Thus RODD forms fundamentally a {\em multiaccess
 channel with erasure}.

Let us also assume that the signaling of each node is subject to unit
average power constraint, i.e., for every $k=1,\dots,K$,
every codeword $(x_{k1},\dots,x_{kM})$
satisfies  $\sum^M_{m=1} s_{km} x^2_{km} \le M$.
 The SNR of the link from node $j$ to node $k$ can
be regarded as $\gamma_{kj} = \gamma_j\, d_{kj}^{-\alpha} |h_{kj}|^2$.
We say node $j$ is a (one-hop) {\em neighbor} or a {\em peer} of node $k$ if
$\gamma_{kj}$ exceeds a given threshold.\footnote{The
 neighbor relationship is not necessarily reciprocal because $\gamma_j
 |h_{kj}|^2$ and $\gamma_k |h_{jk}|^2$ need not be identical.}
Let the set of neighbors (or peers) of $k$
be denoted as $\partial k$, which is also called its {\em
 neighborhood}.  We are only interested in communication over links
between neighbors.  The model~\eqref{eq:MAC} can be reduced to
\begin{equation} \label{eq:MAC1} %
 Y_{km} = (1-s_{km}) \sum_{j\in \partial k} \sqrt{\gamma_j}
 d_{kj}^{-\alpha/2} h_{kj} s_{jm} X_{jm} + V_{km}
\end{equation}
where $V_{km}$ consists of the additive noise $W_{km}$ as well as the
aggregate interference caused by non-neighbors.

Note that~\eqref{eq:MAC} and~\eqref{eq:MAC1} model the half-duplex
constraint at a fundamental level: If node $k$ transmits during a
slot, then its received signal during the slot is erased.

\subsection{A Deterministic Model}

It is instructive to consider a simplification of the preceding
models by assuming noiseless non-coherent reception and energy
detection.  That is, as long as some neighbor transmits energy during
an off-slot of node $k$, a ``1'' is observed in the slot, whereas if
no neighbor emits energy during the slot, a ``0'' is observed.  This can be 
described as an {\em inclusive-or} multiaccess channel (referred to as
OR-channel) with erasure:
\begin{equation} \label{eq:ORMAC} %
 \hat{Y}_{km} = (1-s_{km}) \left( \thevee_{j\in\partial k} (s_{jm} Z_{jm}) \right)
\end{equation}
for $m=1,\dots,M$, where the binary inputs $Z_{jm}$ and outputs
$\hat{Y}_{km}$ take values from $\{0,1\}$.
Since the output is a deterministic function of the
inputs,~\eqref{eq:ORMAC} belongs to the family of {\em deterministic
 models}, which have been found to be a very effective tool in
understanding multiuser channels (see,
e.g,~\cite{ElGCos82IT,AveDig09}).  Despite its simplicity, it captures
the superposition nature of the physical channel, while ignoring the
effect of noise and interference, although those impairments can also
be easily included in the model.

\begin{figure}
 \begin{center}
 \small
 \begin{picture}(0,0)
    \put(15,4){$\ZZZ_1$}
    \put(15,18){$\YYY_1$}
    \put(15,38){$\ZZZ_2$}
    \put(15,53){$\ZZZ_3$}
    \put(15,68){$\ZZZ_4$}
 \end{picture}
 \begin{picture}(245,85)(-24,0)
 \includegraphics[width=3in]{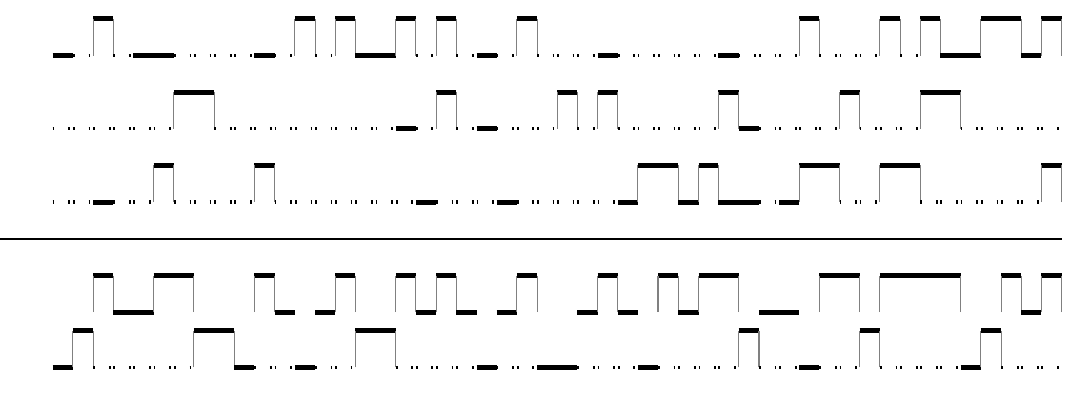}
 \end{picture}
 \end{center}
 \caption{RODD signaling over an OR-channel with erasure.}
\label{f:Superpose}
\end{figure}
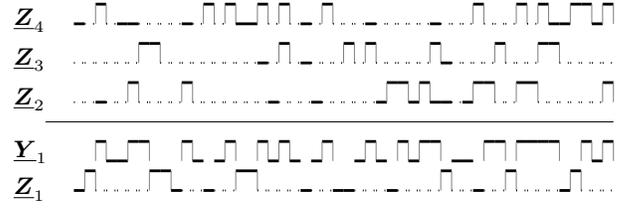

Fig.~\ref{f:Superpose} illustrates a snapshot of RODD signals of four
nodes taken over 50 slots. Here $\ZZZ_1,\dots,\ZZZ_4$ represent the
transmitted signals of node~1 through node~4, respectively, where the
solid lines represent on-slots and the dotted lines represent off-slots.
The received signal of node~1 through its off-slots is $\YYY_1$, which
is the superposition of $\ZZZ_2$, $\ZZZ_3$, and $\ZZZ_4$ with erasures
at its own on-slots (represented by blanks).
Over the period of a single RODD frame, every node can
``simultaneously'' broadcast a message to its neighbors and receive a
message from every neighbor at the same time.

\section{Throughput Results}
\label{s:capacity}

Suppose each node has a message to broadcast to all its (one-hop)
neighbors by transmitting a frame over $M$ slots.
An $M$-slot frame is regarded as being successful for a given node if
its message is decoded correctly by all neighbors; otherwise the frame
is in error.  A rate tuple for the $K$ nodes is achievable if there
exists a code using which the nodes can transmit at their respective
rates with vanishing error probability in the limit where the frame
length $M\to\infty$.

The achievable rates obviously depends on the network topology and the
duplex masks.
It is assumed that every node has complete knowledge of the duplex
masks of all peers (see Section~\ref{s:peer} on neighbor discovery).
For simplicity, in Sections~\ref{s:det} and~\ref{s:gau},
we first consider 
a symmetric network of $K$ nodes who are neighbors of each other,
where the gain between every pair of nodes is identical.  
%
Suppose the elements $s_{km}$ of the duplex masks are
independent identically distributed (i.i.d.) Bernoulli random
variables with $\Prob(s_{km}=1) = q$.

In the simplest scenario, all nodes use randomly generated
i.i.d.~codebooks dependent on the parameters $(K,M,q)$ but independent
of the duplex masks otherwise.  Such a code is called a {\em
  signature-independent code}.
Alternatively, nodes may use {\em signature-dependent codes}, where
the codebooks may depend on the signature pattern
$\S_m=\{s_{1m},s_{2m},\dots,s_{Km}\}$ in every slot $m$. 
Since all nodes are symmetric, dependence on the signatures is only
through the weight of the pattern $\S_m$. 

The amount of information that a node can transmit during a frame is
an increasing function of the number of on-slots,
which in turn has a negative impact on the amount of information it
can collect.
If the amount of information a node receives is several times of the
amount of information it transmits, its signature should consist of
many more off-slots than on-slots, i.e., it is somewhat sparse. 

In case all messages from different nodes are of the same number of
bits, the rate tuple collapses to a single number. The maximum
achievable such rate by using signature-independent
(resp. signature-dependent) codes is called the {\em symmetric rate}
(resp. {\em symmetric capacity}). 

\subsection{The Deterministic Model}
\label{s:det}


Consider the OR-channel described by~\eqref{eq:ORMAC}.
A node's codeword
is basically erased by its own signature mask before transmission.

\begin{proposition}
 The symmetric rate and the symmetric capacity of the OR-channel~\eqref{eq:ORMAC} are
 \begin{align}
    R &= \max_{p \in [0,1]}\frac{1}{K-1}\sum_{n=1}^{K-1}
    \binom{K-1}{n} q^n(1-q)^{K-n} H_2(p^n) \label{eq:RateOR} \\
    C &= \frac{1}{K-1}\big[(1-q)-(1-q)^K\big] \label{eq:CapaOR}
 \end{align}
 where $H_2(p)=-p\log p-(1-p)\log(1-p)$ is the binary entropy
 function.
\end{proposition}

The detailed proof is omitted due to space limitations.
The symmetric rate is achieved by random codebooks with i.i.d.\
Bernoulli$(1-p)$ entries where $p$ maximizes~\eqref{eq:RateOR}. To see
this, consider any given signature pattern 
$\S_m=\s$ in slot $m$, in which $n$ nodes transmit while the remaining
$K-n$ of them listen, the contribution of the slot to the achievable
rate is then given by the mutual information between the binary
received signal $Y$ and the transmitted symbols $\ZZZ$ in the slot:
\begin{align}
 I(\ZZZ;Y|\S_m=\s) &= H(Y|\S_m=\s) - H(Y|\ZZZ,\S_m=\s) \nonumber \\
 &= H(Y|\S_m=\s) \\
 &= H_2(p^n)
\end{align}
where the second equality is due to the deterministic nature of the
model.

The symmetric capacity is higher than the symmetric rate because there
is gain to adapt the codebooks to the signatures.  Basically the
codebook entries at each slot are generated as independent Bernoulli
random variables whose mean value depends on the number of
transmitting nodes in the slot (aka the weight of $\S_m$).  The
parameters of the Bernoulli variables can be optimized for achieving
the capacity.

\begin{figure}
 \begin{center}
\hspace{-25pt}\includegraphics[width=1.1\columnwidth]{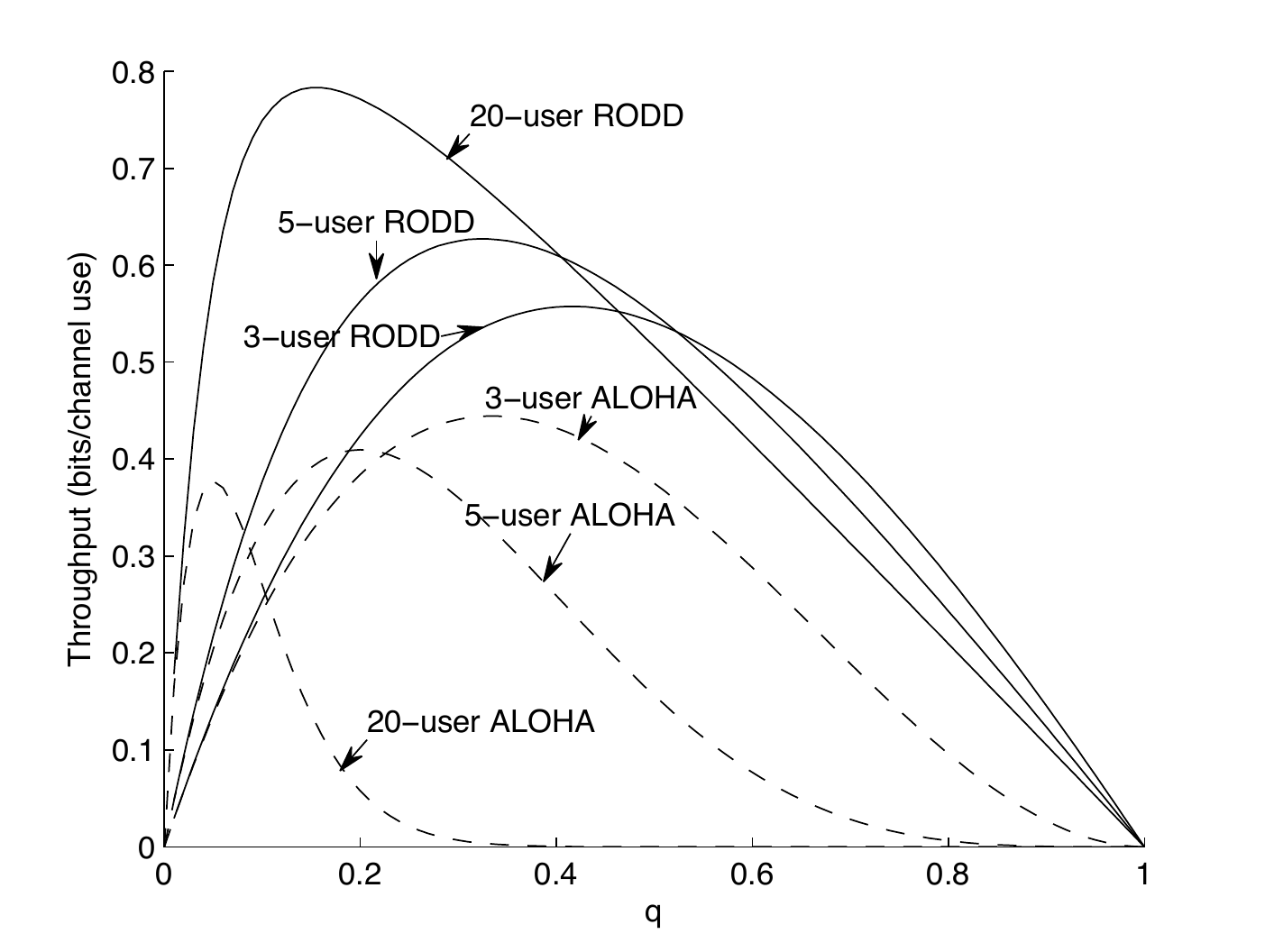}
\end{center}
 \caption{Comparison of the throughput of RODD and ALOHA over
    OR-channel.}
 \label{f:comp}
\end{figure}

We next compare the throughput of a RODD-based scheme with that of ALOHA-type
random access schemes over the same channel~\eqref{eq:ORMAC}, where the throughput is defined as the sum rate of all nodes.
During each frame interval (or contention period), every node in ALOHA
independently chooses either to transmit (with probability $q$) or to
listen (with probability $1-q$) and the choices are independent across
contention periods.  A node successfully broadcasts its message to all
other nodes if the frame is the only transmission during a given frame
interval.  It is easy to see that the throughput of the system with
ALOHA is $Kq(1-q)^{K-1}$, which achieves the maximum
$(1-1/K)^{K-1}$ with $q=1/K$.

For three different node populations ($K=3,5,20$), the comparison
between RODD and ALOHA is shown in Fig.~\ref{f:comp}.
The sum symmetric rate achieved by signature-independent codes is
plotted for RODD.  Clearly, the maximum throughput of RODD is much
higher than that of ALOHA, where the gap increases as the number of
nodes increases.  In fact the throughput of RODD exceeds that of ALOHA
for all values of $q$.  In case of a large number of nodes, the
throughput of ALOHA approaches $1/e$.  On the other hand, with
$p=1-2^{-\frac1{(K-1)q}}$, the total throughput achieved by using RODD
signaling approaches $1-q$ as $K\to\infty$, which is also the
asymptotic sum capacity of RODD achieved by signature-dependent codes. 

The reason for the inferior performance of ALOHA is largely due to
packet retransmissions after collision.
Even if multi-packet reception is allowed,
the throughput of ALOHA is still far inferior compared to RODD
signaling due to the half-duplex constraint.  This is because, in the
case of broadcast traffic studied here, if two nodes simultaneously
and successfully transmit their packets to all other nodes, they still
have to exchange their messages using at least two additional
transmissions.

\subsection{The Gaussian Multiaccess Channel}
\label{s:gau}

\begin{figure}
 \begin{center}
\hspace{-25pt}\includegraphics[width=1.1\columnwidth]{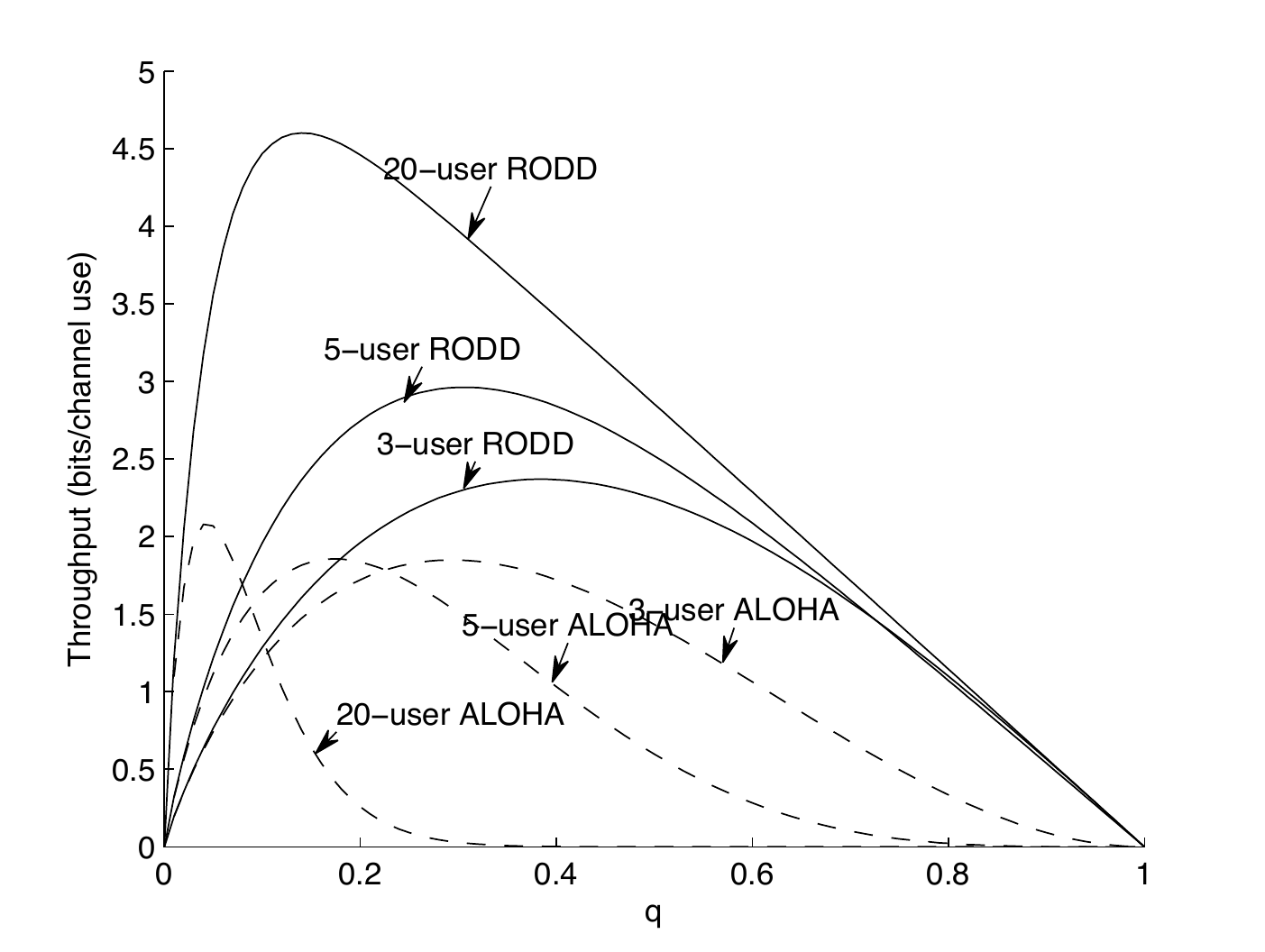}
 \end{center}
 \caption{Comparison of the throughput of RODD and ALOHA over Gaussian
    multiaccess channel at SNR $\gamma=20$ dB.}
 \label{f:Gau}
\end{figure}

Consider now a (non-fading) Gaussian multiaccess channel described
by~\eqref{eq:MAC}, where $d_{kj}=1$, $h_{kj}=1$ for all $k,j$, and
$\{W_{ji}\}$ are i.i.d.\ Gaussian random variables.  
For simplicity, let all nodes be of the same SNR, $\gamma_j=\gamma$.
Let the average power of each transmitted codeword be 1.  Since each
node only transmits over about $qM$ slots, the average SNR during each
active slot is essentially $\gamma/q$.

It is easy to see that the throughput of ALOHA over the Gaussian
channel is $K q(1-q)^{K-1} g(\gamma/q)$, where
$g(x)=\frac{1}{2}\log(1+x)$.
Similar to the results for the deterministic model, we can show that
the symmetric rate and the symmetric capacity for the Gaussian multiaccess channel are achieved with Gaussian codebooks by signature-independent codes and signature-dependent codes, respectively:
\begin{proposition}
 The symmetric rate and the symmetric capacity of the non-fading Gaussian multiaccess channel described by~\eqref{eq:MAC} are
 \begin{align}
    R &= \frac{1}{K-1} \sum_{m=1}^{K-1} \binom{K-1}{m} q^m(1-q)^{K-m}
    g\left(\frac{m\gamma}{q}\right)  \label{eq:RateG} \\
    C &= \frac{1}{K-1} \sum_{m=1}^{K-1} \binom{K-1}{m} q^m(1-q)^{K-m}
    g(w_m) \label{eq:CapaG}
 \end{align}
 where $w_m=\max(\frac{K-m}{K-1}v-1,0)$ and $v$ is chosen to
 satisfy
 \begin{align}   \label{eq:1}
\frac{1}{K}\sum_{m=1}^{K-1}\binom{K}{m}q^m(1-q)^{K-m}w_m=\gamma \ .
\end{align}
\end{proposition}

The case of signature-dependent codes can be regarded as allocating
different powers to different signature patterns in a parallel
Gaussian multiaccess model. 
As is shown in Fig.~\ref{f:Gau}, the throughput of RODD with
signature-independent codes is higher than that of ALOHA
for all number of nodes and every value of $q$.  The more nodes in the
network, the more advantage of RODD signaling.

\subsection{The Achievable Asymmetric Rates}

In many applications, the amount of data different nodes
transmit/broadcast can be very different.  In random access schemes,
nodes with more data will contend for more resources.
The data rate, transmit power and modulation format of a RODD-based
codebook can be adapted to the amount of data to be transmitted.

Suppose the elements $s_{km}$ of node $k$'s signature are i.i.d. Bernoulli random variables with $\Prob(s_{km}=1) = q_k$. Here we study the asymmetric rate region of RODD achieved by signature-independent codes under the fading channel model described in \eqref{eq:MAC}.

\begin{proposition}
The rate tuple $(R_1,\dots,R_K)$ is achievable over the
fading channel~\eqref{eq:MAC} if
\begin{equation}
    R_k \leq \min_{i \neq k}(1-q_i)\sum_{\mathcal{A} \subseteq \mathcal{K}\setminus\{i\}, k \in \mathcal{A}}\frac{\gamma_{ik}}{q_kh_{\mathcal{A}}^i} g(h_{\mathcal{A}}^i)\prod_{j \in \mathcal{A}}q_j\prod_{l \notin \mathcal{A}}(1-q_l) \label{eq:RateFading}
 \end{equation}
for $1 \leq k \leq K$,
 where $\mathcal{K}=\{1,2,\dots,K\}$ and $h_{\mathcal{A}}^i$ is defined as
 $h_{\mathcal{A}}^i=\sum_{j\in \mathcal{A}}\frac{\gamma_{ij}}{q_j}$.
\end{proposition}

Similar as discussed in the case for the Gaussian Multiaccess channel, the asymmetric rate tuple given in \eqref{eq:RateFading} can be achieved by signature-independent codes with random Gaussian codebooks.


\section{Signature Distribution and Peer Discovery}
\label{s:peer}

It is not necessary to directly distribute the set of $K$ duplex masks
to each node in the network.  It suffices to let nodes
generate their signatures using the same pseudo-random number
generator or some other deterministic function
with their respective unique network interface address (NIA)
as the seed.  Every node can in principle reconstruct all
signatures by enumerating all NIAs. 

Before establishing data links, a node needs to acquire the
identities or NIAs of its neighbors.  This is called {\em neighbor discovery}.
Conventional discovery schemes are based on random access, where each
node transmits its NIA many times with random delay, so that
after enough transmissions, every neighbor receives it
at least once without collision~\cite{BorEph07AdhocNet,
  VasTow09ICMCN, NiSri10SigMet}.   As we shall see next,
network-wide full-duplex discovery is achievable using RODD signaling,
where all nodes simultaneously send their (sparse) on-off signatures
and make measurements through their respective off-slots.

The linear multiaccess channel 
model~\eqref{eq:MAC} applies to the neighbor discover problem if
$X_{jm}$, $m=1,\dots,M$, are replaced by the same indicator variable
$B_j$, where $B_j=1$ if node $j$ is present in the neighborhood, and
$B_j=0$ otherwise.  The signal each node $k$ transmits over the entire 
discovery period to identify itself is then the signature $\s_k$ (this
signature need not be the same or of the same length as the one used
for data communication).
Take node 1, for example, whose observation made through its 
off-slots can be expressed (using a simplification of 
model~\eqref{eq:MAC}) as a vector 
\begin{align}
\textstyle  \Y = \sum_{k=2}^K X_k \s_k + \W
\end{align} 
where $X_k$ (which incorporates fading and 
path loss) is zero or is close to zero except if node $k$ is in the
neighborhood of node $1$ and transmits
$\s_k$ during the off-slots of user~1.  The vector $[X_1,\dots,X_K]$ is typically extremely sparse. 

In~\cite{LuoGuo08Allerton, LuoGuo09Allerton}, Luo and Guo have pointed 
out that to identify a small number of neighbors out of a large 
collection of nodes based on the signal received over a linear channel 
is fundamentally a {\em compressed sensing (or sparse recovery)}\, 
problem, for which a small number of measurements (channel uses) 
suffice~\cite{Donoho06IT, CanTao06IT}.\footnote{Several authors have studied
user activity problem in cellular networks using multiuser detection
techniques~\cite{LinLim04TC, AngBig07ITA, AngBig07ICASSP}.  These
works assume channel coefficients are known to the receiver, which
is not the case in most networks.}
Using {\em pseudo-random on-off signatures} for neighbor discovery was
proposed in~\cite{LuoGuo08Allerton, LuoGuo09Allerton} along
with a group testing algorithm.
The key observation
is that, from one node's viewpoint, for each slot with (essentially)
no energy received, all nodes who would have transmitted a pulse
during that slot cannot be a neighbor.
A node basically goes through every off-slot and eliminates nodes
incompatible with the measurement; the surviving nodes are then
regarded as neighbors.  The compressed neighbor discovery scheme
requires only noncoherent energy detection and has been shown to be
effective and efficient at moderate SNRs; moreover, it requires many
fewer symbol transmissions than conventional neighbor discovery
schemes.

With improvement over the algorithms of~\cite{LuoGuo09Allerton},
numerical examples  
show that in a network of $N=\;$10,000  Poisson-distributed nodes, where
each node has on average 50 neighbors, 99\% discovery
accuracy is achieved using 2,500-bit signatures at moderate SNR, 
less than half of that needed by random access discovery to achieve
the same accuracy.  Only one frame of transmission is
needed here, as opposed to many frames in the case of random access,
thus offering significant additional reduction of timing and error-control
overhead embedded in each frame.

Since RODD transmission and neighbor
discovery share the same linear channel model,
it is possible for a new node in a neighborhood to carry out neighbor
discovery solely based on a frame of data transmission by all peers
over the multiaccess channel, without an explicit neighbor
discovery phase.


\section{Channel Coding for RODD}
\label{s:coding}

\subsection{Capacity-achieving Codes}

%
From individual receiver's viewpoint, the channel with RODD is a
multiaccess channel with erasure at known positions.
All good codes for multiaccess
channels are basically good for RODD.  Coding schemes for the OR
multiaccess channel have been studied, e.g., in~\cite{GriCas06ISIT}
and~\cite{GriCas06GC}.  In particular, the nonlinear trellis codes
of~\cite{GriCas06GC} achieves about 60\% of the sum-capacity.


Coding for the Gaussian multiaccess channel is well-understood.  In
particular, Gaussian codebooks achieve the capacity.  In practice,
however, QAM or PSK signaling is often used depending on the SNR.
Practical codes have been shown to be effective in~\cite{SanPel05IT,
 HarRaj09ISIT, BraAul01ICC}.  For example, the codes
of~\cite{SanPel05IT} is based on LDPC codes, where it is pointed out
that degree optimization for the multiuser scenario is important in
this case.  Reference~\cite{BraAul01ICC} is based on trellis-coded
multiple access, which can be particularly suitable for higher
constellations.  There has also been study on rateless codes for
multiaccess channels, e.g.,~\cite{NieEre06GC, UppHos07ISIT}.

Also relevant is a large body of works on channel codes for
code-division multiple access (CDMA).
By
regarding spreading and channel code as the inner and outer codes,
respectively, turbo decoding has been found to be highly effective for
such systems~\cite{RasGra09}.
In the case of RODD, the on-off signatures and individual node's
channel code and can be viewed as inner and outer codes, which
suggests that turbo decoding can be highly effective.

\subsection{A Simple, Short Code Based on Sparse Recovery}

A simple channel code for RODD is proposed in~\cite{GuoZha11Infocom},
which does not achieve the capacity but is simple and efficient if the
messages exchanged between peers consist of a relatively small number
of bits.

As in the peer discovery problem, decoding of this simple code
is essentially via
sparse recovery.  Consider the simplest case, where each node has one
bit to broadcast to all other nodes.  Let node $k$ be assigned
two on-off signatures so that the node
transmits $\s_{k,1}$ to send message ``1,'' and transmits
$\s_{k,0}$ to send message ``0.''  All nodes transmit their signatures
simultaneously, and listen to the channel through their respective
off-slots.  Clearly, this is similar to the neighbor discovery
problem, except that each node tries to identify which signature from
{\em each} neighbor was transmitted so as to recover 1 bit of
information from the node.

The preceding coding scheme can be easily extend to the case where the
message $m_k$ from node $k$ is chosen from a small set of messages
$\{1,\dots,\mu\}$.  In this case, node $k$ is assigned $\mu$ distinct
on-off signatures, and transmit the signature corresponding to its
message.  All signatures are known to all nodes.
The problem is now for each node to identify, out of a
total of $\mu K$ messages (signatures) from all nodes, which $K$ messages
(signatures) were selected.  For example, in case of 10 nodes each with
a message of 10 bits, the problem is to identify $10$ out of
$10\times2^{10}=$10,240 signatures.
A rich set of efficient and effective decoding algorithms are
studied in~\cite{GuoZha11Infocom}.

\section{Synchronization}
\label{s:sync}

In order to decode the messages from the neighboring nodes, it is crucial not
only to acquire their signatures, but also their timing (or relative
delay).  Acquisition of timing is in general a prerequisite to decoding data
regardless of what physical- and MAC-layer technologies are used, thus
RODD is not at a disadvantage compared to other schemes.
In a RODD system, nodes with data may transmit over every frame,
providing abundant cues for timing acquisition and synchronization.
Timing acquisition and decoding are generally easier if the frames
arriving at a receiver are fully synchronous locally within each neighborhood.
To maintain synchronicity in a dynamic network requires extra overhead.
Synchronization is, however, not a necessity. Synchronous or not,
each node collects essentially the same amount of information
through its own off-slots.

\subsection{Synchronous RODD}
Synchronicity has been studied extensively in the context of ad hoc
and sensor networks.  Various distributed algorithms for reaching
consensus~\cite{SchRib08TSP, SchGia08TSP} can be applied to achieve
local synchronicity, e.g., by having each node shift its timing to the
``center of gravity'' of the timings of all nodes in the neighborhood.
The timing still fluctuates over the network, but 
is a smooth function geographically. 
The accuracy of synchronization is limited by two
factors: the channel impairments and the propagation delay.
Synchronization is easy if a RODD slot can be much longer than the
propagation delay across the diameter of a neighborhood.  For example,
a slot interval of 100 $\mu$s would be 100 times the propagation time
of 1 $\mu$s over a 300-meter range.  
For high-rate communication, a RODD symbol can be in the form of an
orthogonal frequency division multiplexing (OFDM) symbol.

Local synchronicity can also be achieved using a common reference,
such as a strong beacon signal.  A possible shortcut is to have all nodes
synchronized using GPS or via listening to base stations in an
existing cellular network, if applicable.  

\subsection{Asynchronous RODD}
In an asynchronous design, the relative delay can be arbitrary, so
that the off-slot of a node is in general not aligned with the on-slot
of its neighbors.  Techniques for decoding asynchronous signals
developed in the context of multiuser
detection/decoding~\cite{Verdu98, CotMul10IT} are generally applicable
to RODD-based systems. 

The algorithms for timing acquisition and synchronization should
account for the fact that an active node can only observe each frame
partially through its off-slots.  To infer about the delays of
neighbors based on partial observations is fundamentally the filtering of a hidden
Markov process.


\section{Concluding Remarks}
\label{s:con}

Proposed and studied in this paper is the novel {\em rapid on-off-division
duplex} scheme, which suggests a radically different design than
conventional FDD and TDD systems.  
On-off signaling has been used since the early days  
of telegraphy, and is also the basis for a simple modulation scheme  
known as on-off keying.  Frequency-hopping multiple-access can be  
regarded as a form of on-off signaling in the frequency domain.  
Recently, transmission through a random on-off mask has been suggested 
to control the amount of interference caused to other nodes in an 
interference channel~\cite{MosKha10ISIT}.  RODD is 
unique in that it exploits on-off signaling to achieve virtual full-duplex 
communication using half-duplex radios.

It is interesting to note that
FDD and TDD suffice in cellular networks because uplink and downlink
transmissions can be assigned regular orthogonal resources.  This absolute
separation of uplink and downlink does not apply to peer-to-peer
networks because, in such a network, one node's downlink resource has
to be matched with its peer's uplink. 
The prevalence of FDD- and TDD-like scheduling schemes
in current ad hoc networks is in part
inherited from the more mature technologies of wired and cellular
networks, and due to the difficulty of separating superposed signals.
Advances in multiuser detection and decoding (e.g.,~\cite{Honig09})
and recent progress in sparse recovery have enabled new technologies
that break away from the model of packet collisions, and hence set the
stage for RODD-based systems.

A rich class of results and techniques in network information theory 
apply to RODD-based systems.  Moreover, almost all technological 
advances in the wireless communications are also applicable to such
systems, including OFDM, multiple antennas, relay, 
cooperation, to name a few. 
In particular, RODD signaling enables virtual full-duplex relaying,
where a relay forwards each received symbol in the next
available on-slot.  The queueing delay
at a relay can be all but eliminated.  This is in contrast to
the store-and-forward scheme used by half-duplex relays, where the
queueing delay at a relay is of the length of one or several frames.

We conclude this paper by describing a specific advantage of RODD for
network state information exchange.
Many advanced wireless transmission techniques require knowledge of
the state of communicating parties, such as the power, modulation
format, beamforming vector, code rate, acknowledgment (ACK), queue length,
etc.  Conventional schemes often treat such {\em network state
information} similarly as data, so that exchange of such information
require a substantial amount of overhead and, in ad hoc networks, 
often many retransmissions.  In a highly mobile network,
the overhead easily dominates the data traffic~\cite{AndJin08CM}. 
By creating a virtual full-duplex channel,
RODD is particularly suitable for nodes to efficiently broadcast
local state information to their respective neighbors.  In fact RODD can 
be deployed as a new sub-layer of the protocol stack, solely devoted
to (virtual full-duplex) state information exchange. 
One potential application of this idea is to assist distributed scheduling by letting
each node choose whether to transmit based on its own state and the states of 
its neighbors. 
We have shown that a simple distributed protocol  lead to an efficient
network-wide TDMA schedule, which typically doubles the throughput of
ALOHA~\cite{HuiGuo10ISIT}. 
Another application is distributed interference management by
exchanging interference prices as studied in~\cite{SchShi09SPM}. 



\section*{Acknowledgment}

The authors would like to thank Martin Haenggi for useful discussions.


\begin{thebibliography}{10}

\bibitem{AndJin08CM}
J.~Andrews, N.~Jindal, M.~Haenggi, R.~Berry, S.~Jafar, D.~Guo, S.~Shakkottai,
  R.~{Heath Jr}, M.~Neely, S.~Weber, A.~Yener, and P.~Stone, ``Rethinking
  information theory for mobile ad hoc networks,'' {\em IEEE Communication
  Magazine}, vol.~46, pp.~94--101, Dec. 2008.

\bibitem{XuSaa01CM}
S.~Xu and T.~Saadawi, ``Does the {IEEE} 802.11 {MAC} protocol work well in
  multihop wireless ad hoc networks?,'' {\em IEEE Communication Magazine},
  vol.~39, pp.~130--137, June 2001.

\bibitem{WinSch00TC}
M.~Z. Win and R.~A. Scholtz, ``Ultra-wide bandwidth time-hopping
  spread-spectrum impulse radio for wireless multiple-access communications,''
  {\em {IEEE} Trans.\ Commun.}, vol.~48, pp.~679 --689, Apr 2000.

\bibitem{AsaSat96IEICE}
H.~Asada, K.~Satou, T.~Yamazato, M.~Katayama, and A.~Ogawa, ``A study on code
  division duplex ({CDD}) for distributed {CDMA} networks,'' {\em Technical
  Report of IEICE}, pp.~89--94, 1996.

\bibitem{Lee02CM}
W.~C.~Y. Lee, ``The most spectrum-efficient duplexing system: {CDD},'' {\em
  IEEE Communication Magazine}, pp.~163--166, 2002.

\bibitem{ShaRap03CM}
S.~Shakkottai, T.~S. Rappaport, and P.~C. Karlsson, ``Cross-layer design for
  wireless networks,'' {\em IEEE Communication Magazine}, vol.~41, pp.~74--80,
  Oct. 2003.

\bibitem{KumRag06AdHocNet}
S.~Kumar, V.~S. Raghavan, and J.~Deng, ``Medium access control protocols for ad
  hoc wireless networks: A survey,'' {\em Ad Hoc Networks}, vol.~4,
  pp.~326--358, May 2006.

\bibitem{RubMor06IFIP}
M.~G. Rubinstein, I.~M. Moraes, M.~Campista, L.~Costa, and O.~Duarte, {\em A
  Survey on Wireless Ad Hoc Networks}, vol.~211 of {\em IFIP International
  Federation for Information Processing}.
\newblock Springer Boston, Nov 2006.

\bibitem{MohKri05}
P.~Mohapatra and S.~Krishnamurthy, {\em {AD HOC NETWORKS: technologies and
  protocols}}.
\newblock Springer Verlag, 2005.

\bibitem{LutHau08ISIT}
T.~Lutz, C.~Hausl, and R.~K\"otter, ``Coding strategies for noise-free relay
  cascades with half-duplex constraint,'' in {\em Proc.\ IEEE Int.\ Symp.\
  Inform. Theory}, pp.~2385--2389, Toronto, ON, Canada, July 2008.

\bibitem{LutKra10ISIT}
T.~Lutz, G.~Kramer, and C.~Hausl, ``Capacity for half-duplex line networks with
  two sources,'' in {\em Proc.\ IEEE Int.\ Symp.\ Inform. Theory},
  pp.~2393--2397, Austin, TX, USA, June 2010.

\bibitem{Kramer07}
G.~Kramer, ``Communication strategies and coding for relaying,'' {\em Wireless
  Networks}, vol.~143 of The IMA Volumes in Mathematics and its Applications,
  pp.~163--175, 2007.

\bibitem{GanGon10ICC}
R.~K. Ganti, Z.~Gong, M.~Haenggi, C.~Lee, S.~Srinivasa, D.~Tisza, S.~Vanka, and
  P.~Vizi, ``Implementation and experimental results of superposition coding on
  software radio,'' in {\em Proc.\ IEEE Int.\ Conf.\ Commun.}, Cape Town, South
  Africa, 2010.

\bibitem{KatGol07Sigcomm}
S.~Katti, S.~Gollakota, and D.~Katabi, ``Embracing wireless interference:
  analog network coding,'' in {\em Proc.\ ACM SIGCOMM}, pp.~397--408, Aug 2007.

\bibitem{HalAnd08Mobicom}
D.~Halperin, T.~Anderson, and D.~Wetherall, ``Taking the sting out of carrier
  sense: interference cancellation for wireless lans,'' in {\em Proc.\ ACM
  Mobicom}, pp.~339--350, Sep 2008.

\bibitem{GolKat08Sigcomm}
S.~Gollakota and D.~Katabi, ``Zigzag decoding: combating hidden terminals in
  wireless networks,'' in {\em Proc.\ ACM SIGCOMM}, pp.~159--170, Aug 2008.

\bibitem{GolPer09Sigcomm}
S.~Gollakota, S.~D. Perli, and D.~Katabi, ``Interference alignment and
  cancellation,'' in {\em Proc.\ ACM SIGCOMM}, pp.~159--170, Aug 2009.

\bibitem{Scholt93Milcom}
R.~A. Scholtz, ``Multiple access with time-hopping impulse modulation,'' in
  {\em Proc.\ IEEE MILCOM}, Bedford, MA, USA, 1993.

\bibitem{WinSch98CL}
M.~Z. Win and R.~A. Scholtz, ``Impulse radio: {H}ow it works,'' {\em {IEEE}
  Commun.\ Lett.}, vol.~2, pp.~36--38, 1998.

\bibitem{ElGCos82IT}
A.~{El Gamal} and M.~H.~M. Costa, ``The capacity region of a class of
  deterministic interference channels (corresp.),'' {\em IEEE Trans.\ Inform.\
  Theory}, vol.~28, pp.~343--346, Mar 1982.

\bibitem{AveDig09}
A.~S. Avestimehr, S.~N. Diggavi, and D.~N.~C. Tse, ``Wireless network
  information flow: A deterministic approach,'' {\em To appear in IEEE Trans.\
  Inform.\ Theory}.
\newblock http://arxiv.org/abs/0906.5394.

\bibitem{BorEph07AdhocNet}
S.~A. Borbash, A.~Ephremides, and M.~J. McGlynn, ``An asynchronous neighbor
  discovery algorithm for wireless sensor networks,'' in {\em Ad Hoc Networks},
  vol.~5, pp.~998--1016, Sep 2007.

\bibitem{VasTow09ICMCN}
S.~Vasudevan, D.~Towsley, D.~Goeckel, and R.~Khalili, ``Neighbor discovery in
  wireless networks and the coupon collector's problem,'' in {\em Proc.~15th
  Annual Int'l Conf.\ Mobile Comput,\ Network..}, pp.~181--192, Beijing, China,
  2009.

\bibitem{NiSri10SigMet}
J.~Ni, R.~Srikant, and X.~Wu, ``Coloring spatial point processes with
  applications to peer discovery in large wireless networks,'' in {\em
  SIGMETRICS}, pp.~167--178, June 2010.

\bibitem{LuoGuo08Allerton}
J.~Luo and D.~Guo, ``Neighbor discovery in wireless ad hoc networks based on
  group testing,'' in {\em Proc.\ Allerton Conf.\ Commun., Control, \&
  Computing}, Monticello, IL, USA, 2008.

\bibitem{LuoGuo09Allerton}
J.~Luo and D.~Guo, ``Compressed neighbor discovery for wireless ad hoc
  networks: the {R}ayleigh fading case,'' in {\em Proc.\ Allerton Conf.\
  Commun., Control, \& Computing}, Monticello, IL, USA, Oct. 2009.

\bibitem{Donoho06IT}
D.~L. Donoho, ``Compressed sensing,'' {\em IEEE Trans.\ Inform.\ Theory},
  vol.~52, pp.~1289--1306, Apr 2006.

\bibitem{CanTao06IT}
E.~J. Candes and T.~Tao, ``Near-optimal signal recovery from random
  projections: Universal encoding strategies?,'' {\em IEEE Trans.\ Inform.\
  Theory}, vol.~52, pp.~5406--5425, Dec. 2006.

\bibitem{LinLim04TC}
D.~D. Lin and T.~J. Lim, ``Subspace-based active user identification for a
  collision-free slotted ad hoc network,'' {\em {IEEE} Trans.\ Commun.},
  vol.~52, pp.~612--621, Apr. 2004.

\bibitem{AngBig07ITA}
D.~Angelosante, E.~Biglieri, and M.~Lops, ``Neighbor discovery in wireless
  networks: A multiuser-detection approach,'' in {\em Proc.\ Inform. Theory
  Appl.\ Workshop}, pp.~46--53, Jan 29-Feb 2 2007.

\bibitem{AngBig07ICASSP}
D.~Angelosante, E.~Biglieri, and M.~Lops, ``A simple algorithm for neighbor
  discovery in wireless networks,'' in {\em Proc. IEEE Int'l Conf. Acoustics,
  Speech and Signal Processing}, vol.~3, pp.~169--172, April 2007.

\bibitem{GriCas06ISIT}
M.~Griot, A.~Casado, W.-Y. Weng, H.~Chan, J.~Basak, E.~Yablonovitch,
  I.~Verbauwhede, B.~Jalali, and R.~Wesel, ``Trellis codes with low ones
  density for the {OR} multiple access channel,'' in {\em Proc.\ IEEE Int.\
  Symp.\ Inform. Theory}, pp.~1817--1821, July 2006.

\bibitem{GriCas06GC}
M.~Griot, A.~Vila~Casado, and R.~Wesel, ``Non-linear turbo codes for
  interleaver-division multiple access on the {OR} channel,'' in {\em Proc.\
  IEEE GLOBECOM}, pp.~1--6, Nov 27-Dec 1 2006.

\bibitem{SanPel05IT}
A.~Sanderovich, M.~Peleg, and S.~Shamai, ``{LDPC} coded {MIMO} multiple access
  with iterative joint decoding,'' {\em IEEE Trans.\ Inform.\ Theory}, vol.~51,
  pp.~1437--1450, Apr 2005.

\bibitem{HarRaj09ISIT}
J.~Harshan and B.~S. Rajan, ``Coding for two-user {G}aussian {MAC} with {PSK}
  and {PAM} signal sets,'' in {\em Proc.\ IEEE Int.\ Symp.\ Inform. Theory},
  pp.~1859--1863, June 28-July 3 2009.

\bibitem{BraAul01ICC}
F.~Brannstrom, T.~Aulin, and L.~Rasmussen, ``Iterative multi-user detection of
  trellis code multiple access using a posteriori probabilities,'' in {\em
  Proc.\ IEEE Int.\ Conf.\ Commun.}, vol.~1, pp.~11--15, June 2001.

\bibitem{NieEre06GC}
U.~Niesen, U.~Erez, D.~Shah, and G.~Wornell, ``Rateless codes for the
  {G}aussian multiple access channel,'' in {\em Proc.\ IEEE GLOBECOM},
  pp.~1--5, Nov 27-Dec 1 2006.

\bibitem{UppHos07ISIT}
M.~Uppal, A.~Host-Madsen, and Z.~Xiong, ``Practical rateless cooperation in
  multiple access channels using multiplexed raptor codes,'' in {\em Proc.\
  IEEE Int.\ Symp.\ Inform. Theory}, pp.~671--675, June 2007.

\bibitem{RasGra09}
L.~K. Rasmussen and A.~Grant, ``Iterative techniques,'' in {\em Advances in
  Multiuser Detection} (M.~Honig, ed.), ch.~2, Wiley-IEEE Press, 2009.

\bibitem{GuoZha11Infocom}
D.~Guo and L.~Zhang, ``Peer-to-peer broadcast in wireless networks via sparse
  recovery,'' {\em submitted to Proc.\ IEEE INFOCOM}.

\bibitem{SchRib08TSP}
I.~D. Schizas, A.~Ribeiro, and G.~B. Giannakis, ``{Consensus in ad hoc WSNs
  with noisy links-part I: Distributed estimation of deterministic signals},''
  {\em IEEE Trans.\ Signal Process.}, vol.~56, no.~1, pp.~350--364, 2008.

\bibitem{SchGia08TSP}
I.~D. Schizas, G.~B. Giannakis, S.~I. Roumeliotis, and A.~Ribeiro, ``{Consensus
  in ad hoc WSNs with noisy links-part II: Distributed estimation and smoothing
  of random signals},'' {\em IEEE Trans.\ Signal Process.}, vol.~56, no.~4,
  pp.~1650--1666, 2008.

\bibitem{Verdu98}
S.~Verd{\'u}, {\em Multiuser Detection}.
\newblock Cambridge University Press, 1998.

\bibitem{CotMul10IT}
L.~Cottatellucci, R.~R. M{\"u}ller, and M.~Debbah, ``{Asynchronous CDMA Systems
  with Random Spreading--Part II: Design Criteria},'' {\em IEEE Trans.\
  Inform.\ Theory}, vol.~56, pp.~1498--1520, 2010.

\bibitem{MosKha10ISIT}
K.~Moshksar and A.~K. Khandani, ``On the achievable rates in decentralized
  networks with randomized masking,'' in {\em Proc.\ IEEE Int.\ Symp.\ Inform.
  Theory}, pp.~420--424, Austin, TX, USA, 2010.

\bibitem{Honig09}
M.~L. Honig, ed., {\em Advances in Multiuser Detection}.
\newblock Wiley-IEEE Press, 2009.

\bibitem{HuiGuo10ISIT}
K.~H. Hui, D.~Guo, and R.~A. Berry, ``Medium access control via nearest
  neighbor interactions for regular wireless networks,'' in {\em Proc.\ IEEE
  Int.\ Symp.\ Inform. Theory}, Austin, TX, USA, 2010.

\bibitem{SchShi09SPM}
D.~A. Schmidt, C.~Shi, R.~A. Berry, M.~L. Honig, and W.~Utschick, ``Pricing
  algorithms for power control and beamformer design in interference
  networks,'' {\em {IEEE} Signal Processing Mag.}, vol.~26, pp.~53--63, 2009.

\end{thebibliography}
\end{document}